\documentclass{elsart}

 \usepackage{graphicx}
\usepackage{epsfig}

\usepackage{amssymb}

\begin{document}
\begin{frontmatter}
\title{Dynamics of coupled Josephson junctions under the influence of applied fields}
\author{Chitra R Nayak}
\ead{rchitra@cusat.ac.in}
and
\author{V C Kuriakose}
\ead{vck@cusat.ac.in}
%\corauth[cor1]{V C Kuriakose}
\address{Department of Physics, Cochin University of Science and Technology, Kochi-682022, India}

%\author{Chitra R Nayak }
%\email{rchitra@cusat.ac.in}
%\author{V C Kuriakose}
%\email{vck@cusat.ac.in} %
%\affiliation{Department of Physics, Cochin
%University of Science and
%Technology, Kochi, 682022} %

%%%%%%%%%%%%

\begin{abstract}
We investigate the effect of the phase difference of applied fields on the
dynamics of mutually coupled Josephson junction. %The effect of an
%applied phase difference on a mutually coupled Josephson junction is
%studied.
 The system desynchronizes for any value of applied phase
difference and the dynamics even changes from chaotic to periodic
motion for some values of applied phase difference. We report that
by keeping the value of phase difference as $\pi$, the system
continues to be in periodic motion for a wide range of system
parameter values which might be of great practical applications.
\end{abstract}%

%\pacs 05.45+b
\end{frontmatter}

\section{Introduction}
The study of the dynamics of Josephson junction (JJ) is of
fundamental and experimental interest. The interaction of
Josephson junctions with external fields have played an
important role in the development of Josephson physics and its
chaotic dynamics \cite{jensen,cron,wen,yeh}. The existence of chaos
in rf-biased Josephson junction has been verified through theory,
numerical simulation  and experiments \cite{kau}. The rf-biased
junction is of practical importance because of its application as
voltage standards, a situation where chaotic behavior is least
desired \cite{rlk}. Control of chaos is an active area of research
\cite{kap} because of the many undesirable effects chaos brings in
mechanical systems and other devices. It was shown to be possible to
control chaos both theoretically and experimentally using different
methods such as giving a feed back \cite{sing}, application of a
weak periodic force \cite{brai} etc. The problem of controlling
spatiotemporal chaotic pattern induced by an applied rf signal in a
Josephson junction has earlier been discussed \cite{olsen}. By
controlling chaos in rf-biased Josephson junctions it was shown that
even in the presence of thermal noise they can be used as voltage
standards \cite{abra}. Suppression of temporal and spatio-temporal
chaos allows complex systems to be operated in highly nonlinear
regimes. This is a desirable feature in many physical systems. By
applying a small time-dependent modulation to a parameter, a chaotic
system can be stabilzed. However in practical applications this
method requires that the characteristic times of the system is not
too short compared with the times of the feed back system. In the
case of JJ oscillators the characteristic times of the dynamics
response are of the orders of few picoseconds which is too short for
any electronic feedback control system.

Since it was shown that chaotic systems could be synchronized by
linking them with common signal \cite{pecora} many works have been
done in this direction because of its application in secure
communication \cite{kim}. Synchronization in Josephson junction has
been an interesting area of research \cite{grib,blackburn,ahmet}. The role
of phase difference of applied sinusoidal fields in desynchronizing
and suppressing chaos in duffing oscillators has been studied
\cite{zhi,yin}. In the present work, we consider the effect of phase
difference of the applied rf-fields on mutually coupled Josephson
junctions. We discuss the equation of coupled JJ in section II and
arrive at the dimensionless first order form of the equation of
motion of the system.  In section III the parameter range in which
the system is synchronized is found  and the effect of phase
difference of the applied rf fields on synchronization is found. We
analyze the dynamics of the Josephson junction after applying the
phase difference. Then by fixing a particular phase value where the
system becomes periodic we discuss the effect of other parameters on
the junction. In section IV the results are discussed and the
applications are mentioned.
%%%%%%%%%%
\section{THEORY}
\begin{figure}[tbh]
\centering
\includegraphics[width=0.8\columnwidth]{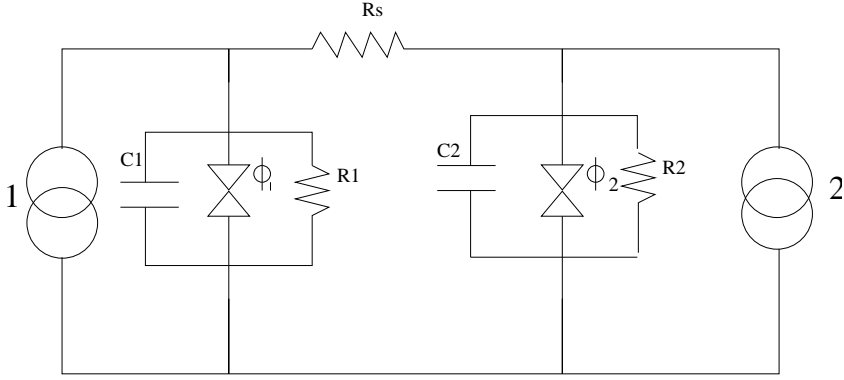}
\caption{Schematic representation of a coupled Josephson junction
connected in parallel with a linking resistor $R_s$. 1 and 2 represents the applied fields.} %
\label{cir}
\end{figure}

Josephson junction can be represented by a resistively and
capacitively shunted junction (RCSJ) model and the dynamics of the
system can be explored by writing the equation of motion \cite{bar}.
The equation of a single Josephson junction by this model can be
written by solving Kirchoff's law as
\begin{equation}
\frac {\hbar C} {2e}\frac {d^2 \phi} {dt^2} + \frac {\hbar}
{2eR}\frac {d \phi} {dt}+ i_{c}\sin\phi=
i_{dc}+i_0\cos(\omega t),%%%
 \label{single}
\end{equation}

where $\phi$ is the phase difference of the wave function across the
junction, $i_0\cos(\omega t)$ is the driving rf - field and $i_{dc}$
is the dc bias. The junction is characterized by a critical current
$i_c$, capacitance $C$ and normal resistance $R$. The coupled JJ
considered here consists of a pair of such junctions wired in
parallel with a linking resistor $R_s$ \cite{blackburn}. Such a
system can be schematically given as in Fig(\ref{cir}) and the
dynamical equations can be written as
\begin{equation}
\frac {\hbar C_1} {2e}\frac {d^2 \phi_1} {dt'^2} + \frac {\hbar}
{2eR_1}\frac {d \phi_1} {dt'}+ i_{c1}\sin\phi_1=
i_{dc}'+i_0'\cos(\omega t')-i_s \label{one}
\end{equation}
\begin{equation}
\frac {\hbar C_2} {2e}\frac {d^2 \phi_2} {dt'^2} + \frac {\hbar}
{2eR_2}\frac {d \phi_2} {dt'}+ i_{c2}\sin\phi_2=
i_{dc}'+i_0'\cos(\omega t'+\theta)+i_s \label{two},
\end{equation}
where $i_s$ is the current flowing through the coupling resistor and
is given as
\begin{equation}
i_s= \frac{\hbar}{2eR_s} \left[\frac{d \phi_1}{dt'}- \frac{d
\phi_2}{dt'}\right]. 
\label{aaaa}
\end{equation}
In order to express Eqs.(\ref{one}) and (\ref{two}) in
dimensionless, normalized form we introduce the junction plasma
frequencies $\omega_{J1}$ and $\omega_{J2}$ which are given by
$\omega_{J1}=\left(2ei_{c1}/\hbar C_{1}\right)^{\frac{1}{2}}$ and
$\omega_{J2}=\left(2ei_{c2}/\hbar C_{2}\right)^{\frac{1}{2}}$ and
also normalized time scale as $t=\omega_{J1} t'$. The dimensionless
damping parameter is defined as
$$\beta=\frac{1}{R_{1}}\sqrt{\frac{\hbar}{2 e i_{c1} C_{1}}}.$$
The dc bias current $i_{dc}'$ and the rf amplitude $i_{0}'$ are
normalized to the critical current $i_{c1}$. We also normalize the
actual frequency $\omega$ to $\Omega=\omega /\omega_{J1}$ and
$\alpha_{s}=\left(R_1 / R_s\right)\beta$.
%%%%%%%%%%%%%%%
\begin{figure}[tbh]
\centering
\includegraphics[width=0.8\columnwidth]{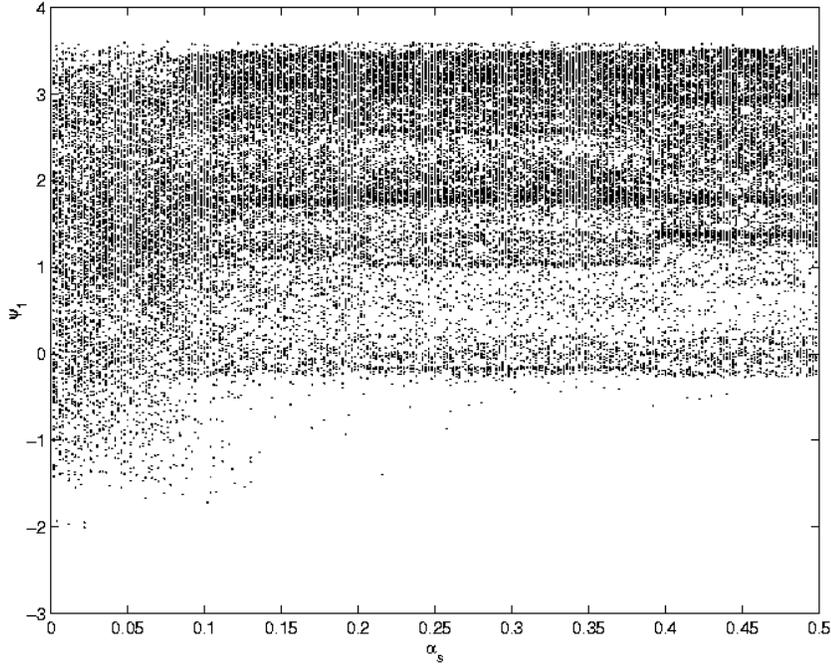}
\caption{Maxima of the normalized voltage plotted against coupling
strength $\alpha_s$ for
$\theta=0$, $\beta=0.15$, $i_0=0.7$ and $\omega=0.6$.} %
\label{vol}
\end{figure}
%%%%%%%%%%%%%%
%%%%%%%%%%%%
\begin{figure}[tbh]
\centering
\includegraphics[width=0.8\columnwidth]{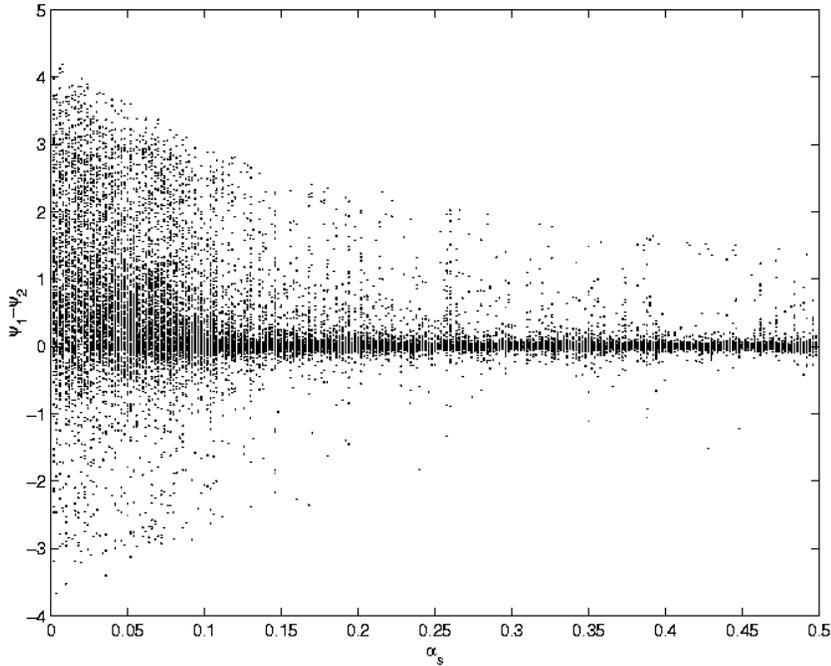}
\caption{Maxima of the difference in voltage against coupling
strength $\alpha_s$. %\newline
$\theta=0$, $ \beta=0.15$, $i_0=0.7$ and $\omega=0.6$.} %
\label{difvol2}
\end{figure}
%%%%%%%%%%
For identical Josephson junctions, we can write Eqs.(\ref{one}) and
(\ref{two}) as
\begin{eqnarray}\nonumber
\ddot{\phi_1}+ \beta \dot{\phi_1}+ \sin\phi_1&=&i_{dc}+i_0
\cos(\Omega
t) - \alpha_s\left[\dot{\phi_1}-\dot{\phi_2}\right],\\
\ddot{\phi_2}+ \beta \dot{\phi_2}+ \sin\phi_2&=&i_{dc}+i_0
\cos(\Omega t+\theta) -
\alpha_s\left[\dot{\phi_2}-\dot{\phi_1}\right].
\end{eqnarray}
It can be seen that the coupling arises as a natural consequence of
the exchange of current through the resistor $R_s$  and it depends
on the differential voltage ($\psi_1-\psi_2$). For Josephson
junction devices, phase derivatives are of central importance
because they are proportional to junction voltages. In order to
study the system numerically we write the above equations in the
first order differential form as
\begin{eqnarray}\nonumber
\label{runge} %
\dot{\phi_1} &=& \psi_1,\\\nonumber %
\dot{\psi_1}&=&-\beta \psi_1-\sin\phi_1+i_{dc}+i_0 \cos(\Omega t)
-\alpha_s \left[\psi_1-\psi_2\right],\\
\dot{\phi_2} &=& \psi_2,\\\nonumber %
\dot{\psi_2}&=&-\beta \psi_2-\sin\phi_2+i_{dc}+i_0 \cos[(\Omega
t)+\theta]- \alpha_s \left[\psi_2-\psi_1\right].
\end{eqnarray}
 Eq.~(\ref{runge}) is studied using fourth order Runge-Kutta method and
the maxima of normalized voltage values are plotted to study the
dynamics. It was observed that the system exhibits chaotic behavior
for $ \beta=0.15$, $i_0=0.7$, $i_{dc}=0.3$ and $\Omega=0.6$ as seen
in Fig.~\ref{vol}. From Fig.~\ref{difvol2} we can see that the
difference in voltage becomes smaller as the coupling strength is
increased.
%Fixing the values of other parameters as $
%\beta=0.15$, $i_0=0.7$ and $\omega=0.6$ we find that the system
%seems to be perfectly synchronized as shown in Fig.~\ref{syn}(a).
 %%%%%%%%%%%

\section{The effect of Phase Difference}
\begin{figure}[tbh]
\centering
\includegraphics[width=0.9\columnwidth]{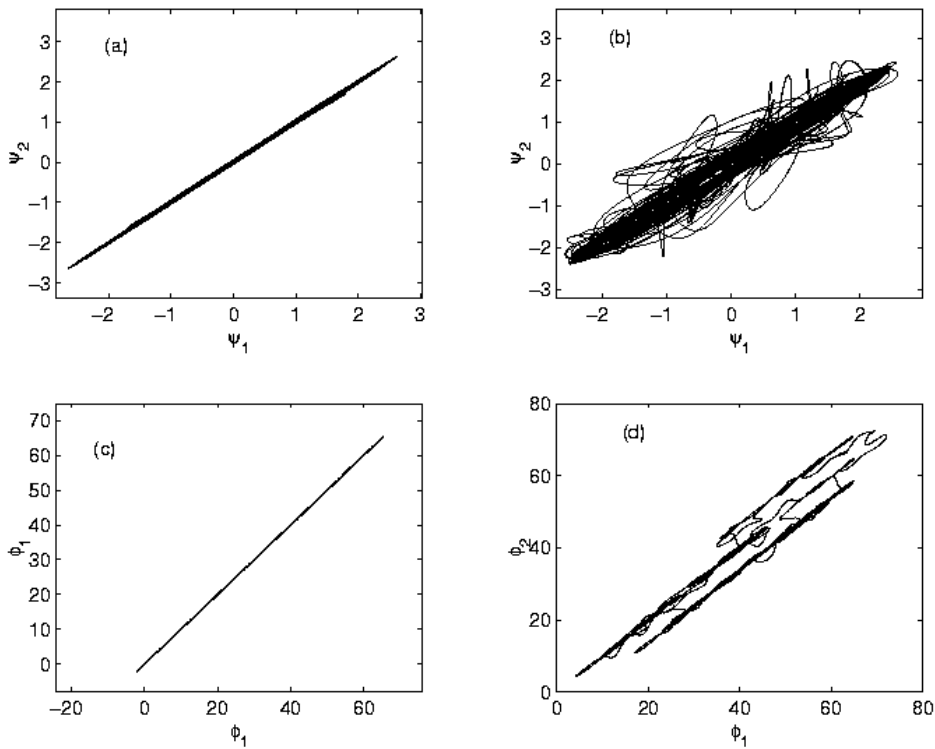}
\caption{a and c show the system is synchronized for $\theta=0$, $
\beta=0.15$, $i_0=0.7$, $i_{dc}=0.3$, $\alpha_s=0.45$ and
$\omega=0.6$.
b and d show the system desynchronized for a phase difference of $\theta=0.1 \pi$} %
\label{syn}
\end{figure}

\begin{figure}[tbh]
\centering
\includegraphics[width=0.9\columnwidth]{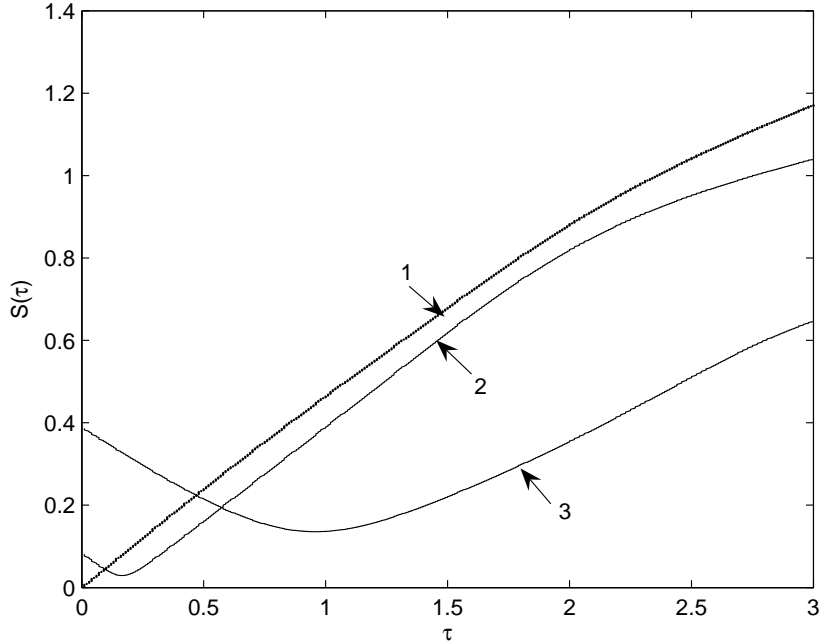}
\caption{Similarity function $S(\tau)$ versus $\tau$ for
different values of phase difference $\theta$. Curve 1 is with phase difference $\theta=0$, 2
for $\theta=0.1\pi$ and 3 for $\theta=0.5\pi$.} %
\label{sim}
\end{figure}

The effect of phase difference of the applied sinusoidal driving
field on coupled periodically driven duffing oscillator has been
studied \cite{yin} and it was observed that the system
desynchronizes by the application of a phase difference. To study
how the phase difference of the applied fields acts on JJ system we
write $S_\phi=\phi_1-\phi_2$ and $S_\psi=\psi_1-\psi_2$ and from
Eq.~(\ref{runge}) we get
\begin{eqnarray}\label{synchro}
\dot{S_\phi}&=& S_\psi, \\\nonumber %
\dot{S_\psi}&=& -\beta S_\psi - \sin\phi_1 + \sin\phi_2 - 2 \alpha_s
S_\psi + 2 i_0 \sin \left(\Omega
t+\frac{\theta}{2} \right)\sin \left(\frac{\theta}{2}\right). %
\end{eqnarray}
When $\theta=0$ the term $2 i_0 \sin \left(\Omega
t+ \theta / 2  \right)\sin \left( \theta / 2\right)$ vanishes. We choose the values of
$\alpha_s$ such that the difference in voltage is negligible. ie.,
$\psi_1 \approx \psi_2$. Now both $\dot{S_\phi}$ and $\dot{S_\psi}$
go to zero. From Fig.~\ref{difvol2} we select our value of
$\alpha_s$ as $0.45$ which almost satisfies this condition. From
Eq.~\ref{synchro} we can see that even for small values of phase
differences $\dot{S_\psi}\neq 0$ and the system desynchronizes.
Figs.~\ref{syn}(a) and \ref{syn}(c) show that the system is
synchronized and Figs.~\ref{syn}(b) and \ref{syn}(d) show that the
system is desynchronized by an applied phase difference of
$\theta=0.1 \pi$. The level of mismatch of chaos synchronization can
be given quantitatively by taking the similarity function $S(\tau)$
as a time averaged difference between the variables $\psi_1$ and
$\psi_1$ taken with time shift $\tau$ \cite{kurth}
\begin{equation}
S^2(\tau)=\frac{\langle\left[\psi_1(t+\tau)-\psi_2(t)\right]^2\rangle}
{\left[\langle\psi_1^2(t)\rangle\right]\left[\langle\psi_2^2(t)\rangle\right]^{1/2}}.
\end{equation}
We plot $S(\tau)$ against $\tau$ for different values of phase
difference $\theta$ as shown in Fig. \ref{sim}. It is observed that for
$\theta=0$, the system is in complete synchronization. For a finite
value of phase difference we observe that a minimum of $S(\tau_0)$
appears which indicates the existence of a certain phase difference
between the interacting systems. $S(\tau_0)$ is finite in these
cases which means that in this regime the amplitudes are
uncorrelated.

\begin{figure}[tbh]
\centering
\includegraphics[width=0.9\columnwidth]{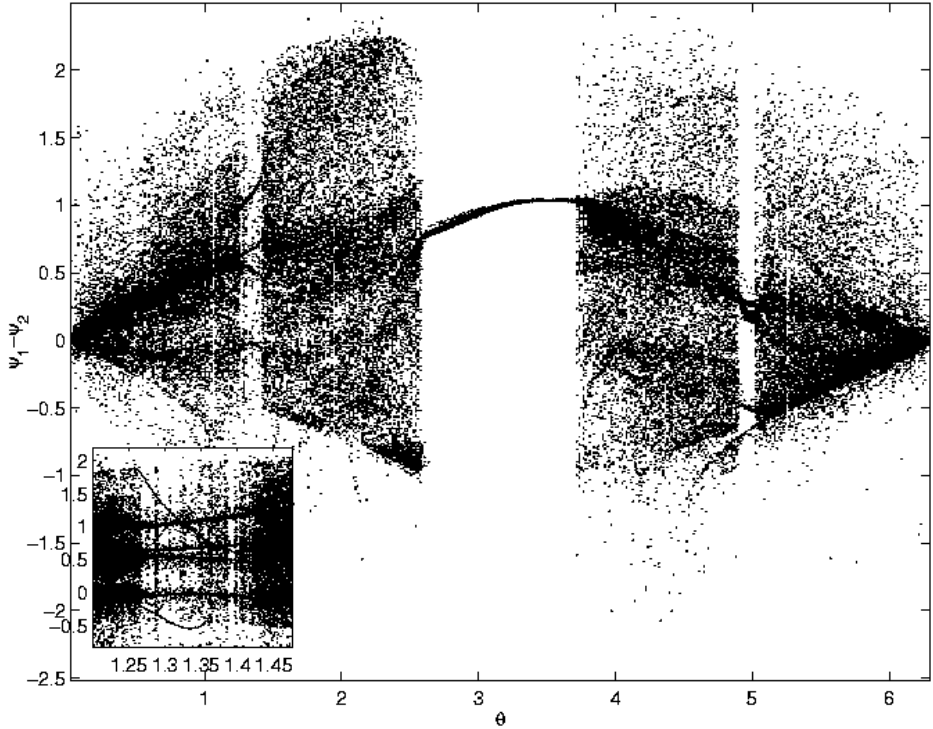}
\caption{Maxima of the  difference in voltage is plotted against
phase difference applied, $\theta=0-2\pi$. $\alpha_s=0.45$,
$i_{dc}=0.3$,
$ \beta=0.15$, $i_0=0.7$ and $\omega=0.6$.}%
\label{difvoltheta}
\end{figure}

Now we vary the values of $\theta$ from $0$ to $2\pi$ and plot the
maxima of the difference in voltage against the phase. It can be
observed from Fig.~\ref{difvoltheta} that the system passes from
chaotic to periodic motion for some values of $\theta$. In the
synchronization manifold, ie., when $\phi_1 = \phi_2$ and $\psi_1
=\psi_2$ we can write $P_\phi=\left[\phi_1+\phi_2\right]/2$ and
$P_\psi=\left[\psi_1+\psi_2\right]/2$ and from Eq.~(\ref{runge}) we
get
\begin{eqnarray}\label{period}
\dot{P_\phi}&=& P_\psi, \\\nonumber %
\dot{P_\psi}&=& -\beta P_\psi -\sin(P_\phi) +idc +  i_0 \cos
\left(\frac{\theta}{2}\right) \cos \left(\Omega
t+\frac{\theta}{2} \right). %
\end{eqnarray}
Eq.~(\ref{period}) is equivalent to Eq.~(\ref{single}) and we can
see that $i_0$ is replaced by $i_0 \cos \left(\theta /2\right)$ and
the phase of the driving field leads by $\theta/2$ as a result of
coupling. This might correspond to a parameter space where the
system is periodic which explains the change in system dynamics.

\begin{figure}[tbh]
\centering
\includegraphics[width=0.9\columnwidth]{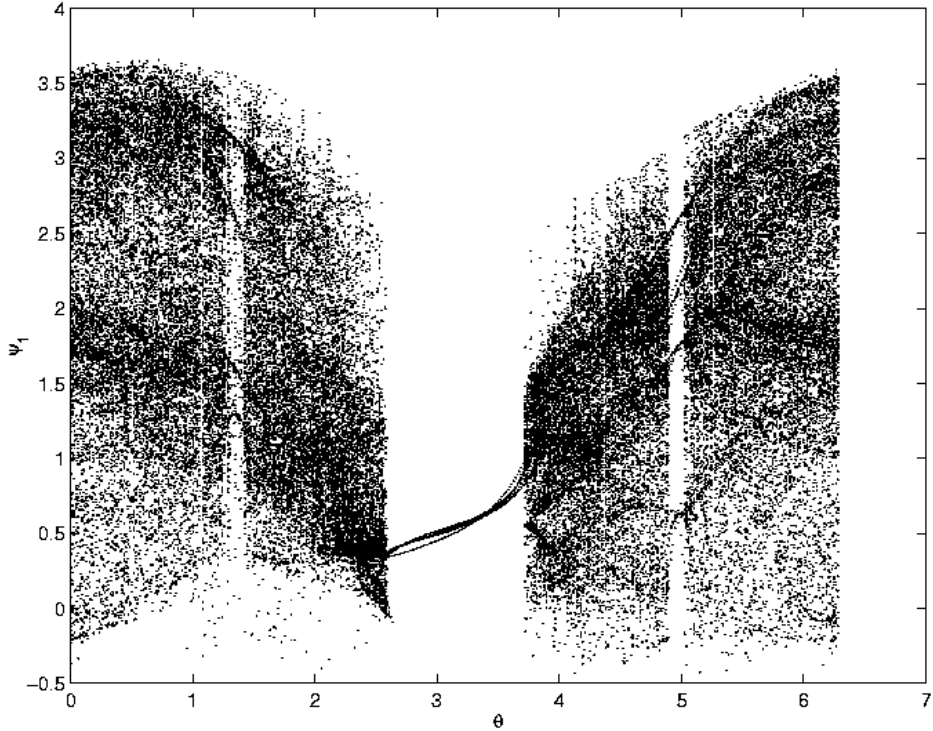}
\caption{Voltage corresponding to a single junction is plotted for
phase values $\theta=0-2\pi$ with $\alpha_s=0.45$, $i_{dc}=0.3$, ,
$ \beta=0.15$, $i_0=0.7$ and $\omega=0.6$.} %
\label{voltheta}
\end{figure}

\begin{figure}[tbh]
\centering
\includegraphics[width=0.9\columnwidth]{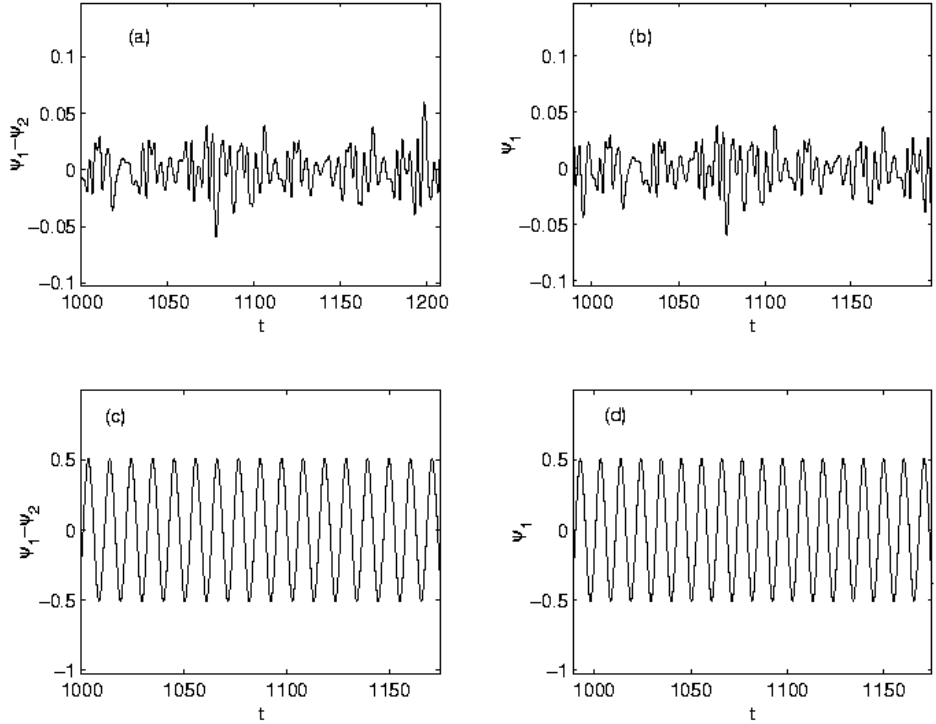}
\caption{(a) shows the differential voltage plotted against time and
(b) is the voltage of one junction with $\theta=0$. It is observed
that the variation is chaotic and the maximum difference in voltage
is $0.05$. In (c) and (d) shows the corresponding voltages with an
applied phase difference of $\pi$.  Other parameter values are
$\alpha_s=0.45$, $\beta=0.15$, $i_0=0.7$ and $\omega=0.6$. }%
\label{difvol}
\end{figure}

 The variation of the voltage of one junction  with the
applied phase difference is shown in Fig.~\ref{voltheta}. From the
inset of Fig.~\ref{difvoltheta} it can be seen that the system
exhibits periodic window in the region where $\theta=0.34 \pi - 0.4
\pi$. However in this region even a slight change in system
parameter values would bring the system back to chaotic regime. For
 a phase difference of $\theta=0.95 \pi-1.5 \pi$ the system exhibits
periodic motion and the corresponding periodic difference in voltage
and voltage plotted against time is shown in Figs.~\ref{difvol}(c)
and \ref{difvol}(d). Figs.~\ref{difvol}(a) and \ref{difvol}(b) shows
difference in voltage and voltage against time for a phase
difference of $\theta=0$.
%%%%%%%%%
%%%%%%%%%%%%
%%%%%%%%%%%%%%%%%%%%%%
%Around a phase difference of  $0.3 \pi$ we get a periodic motion as
%seen from the inset of  Fig. \ref{difvoltheta}.
%However in this region, even a small change in any parameter value
%will take the system from periodic to chaotic state.
%%%%%%%%%%%%%%%%%%%%%%

\begin{figure}[tbh]
\centering
\includegraphics[width=0.8\columnwidth]{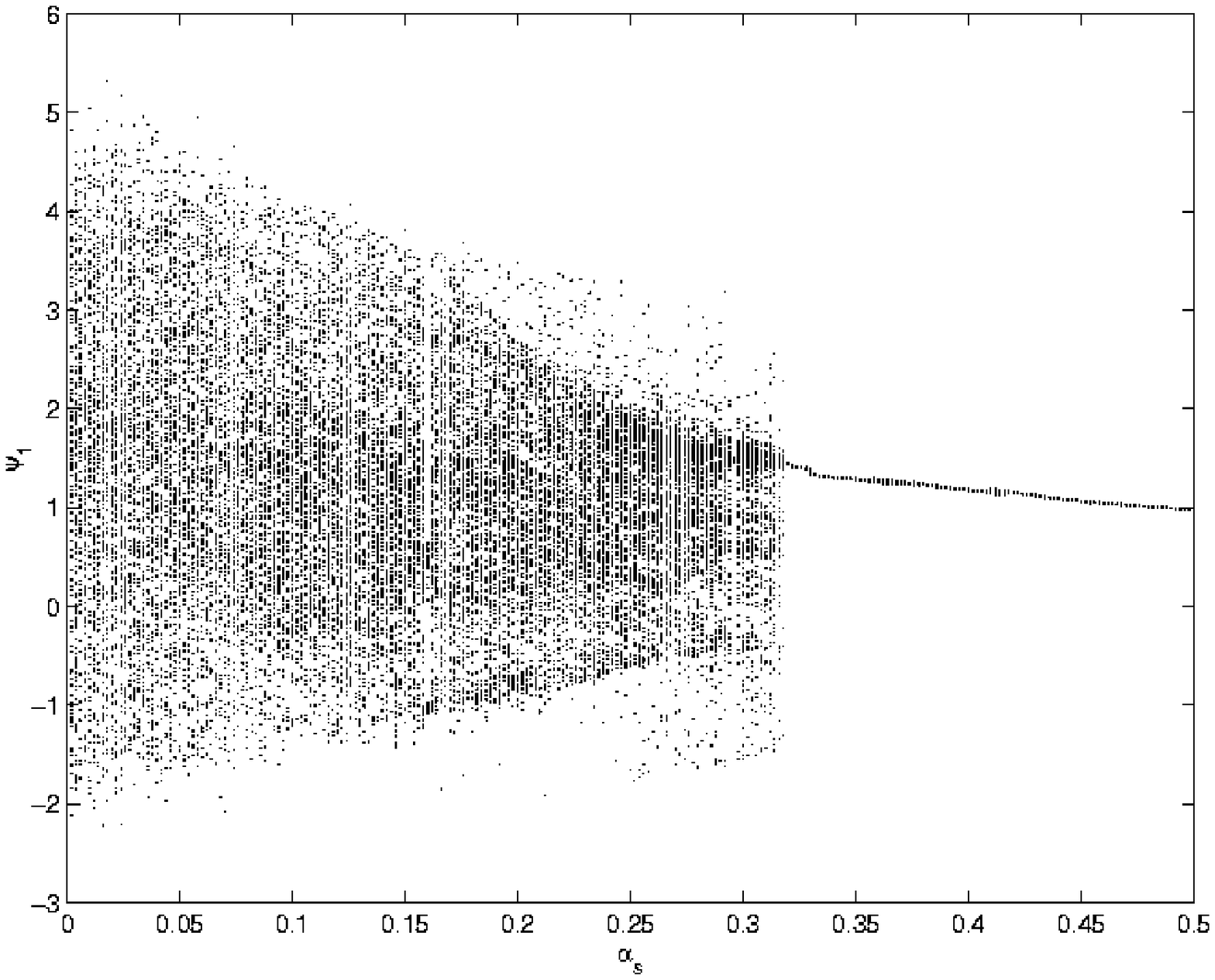}
\caption{Maxima of the normalized voltage against coupling strength
$\alpha_s$.
$\theta=\pi$, $ \beta=0.15$, $i_{dc}=0.3$, $i_0=0.7$ and $\omega=0.6$.} %
\label{volcouthe}
\end{figure}
%%%%%%%%%%%%%%%%%%
Fixing the phase difference between the driving fields as $\pi$, the
change in the response of the system to other parameter variations
are studied. From Fig.~\ref{vol} we can see that the system is in
chaotic motion even for large values of coupling strength. However
by the application of a phase difference of $\pi$ we observe that
the system exhibits periodic motion from a coupling strength $0.32$
onwards as can be seen from Fig.~\ref{volcouthe}.

\begin{figure}[tbh]
\centering
\includegraphics[width=0.8\columnwidth]{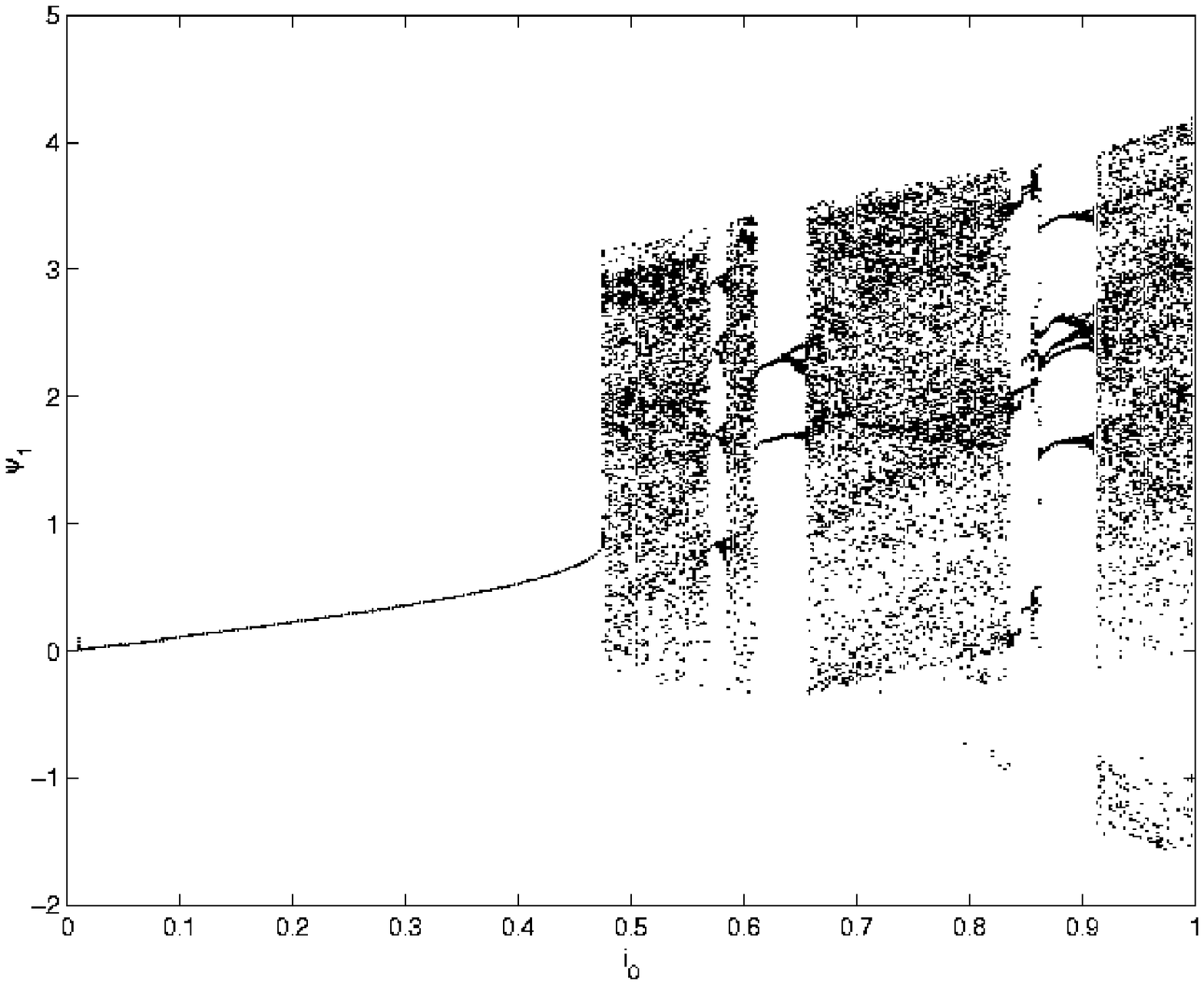}
\caption{Maxima of the normalized voltage against amplitude of
applied field $i_0$ with $\theta=0$. The other parameter values are
$\alpha_s=0.45$, $i_{dc}=0.3$,
$ \beta=0.15$ and $\omega=0.6$.} %
\label{volepsi}
\end{figure}

\begin{figure}[tbh]
\centering
\includegraphics[width=0.8\columnwidth]{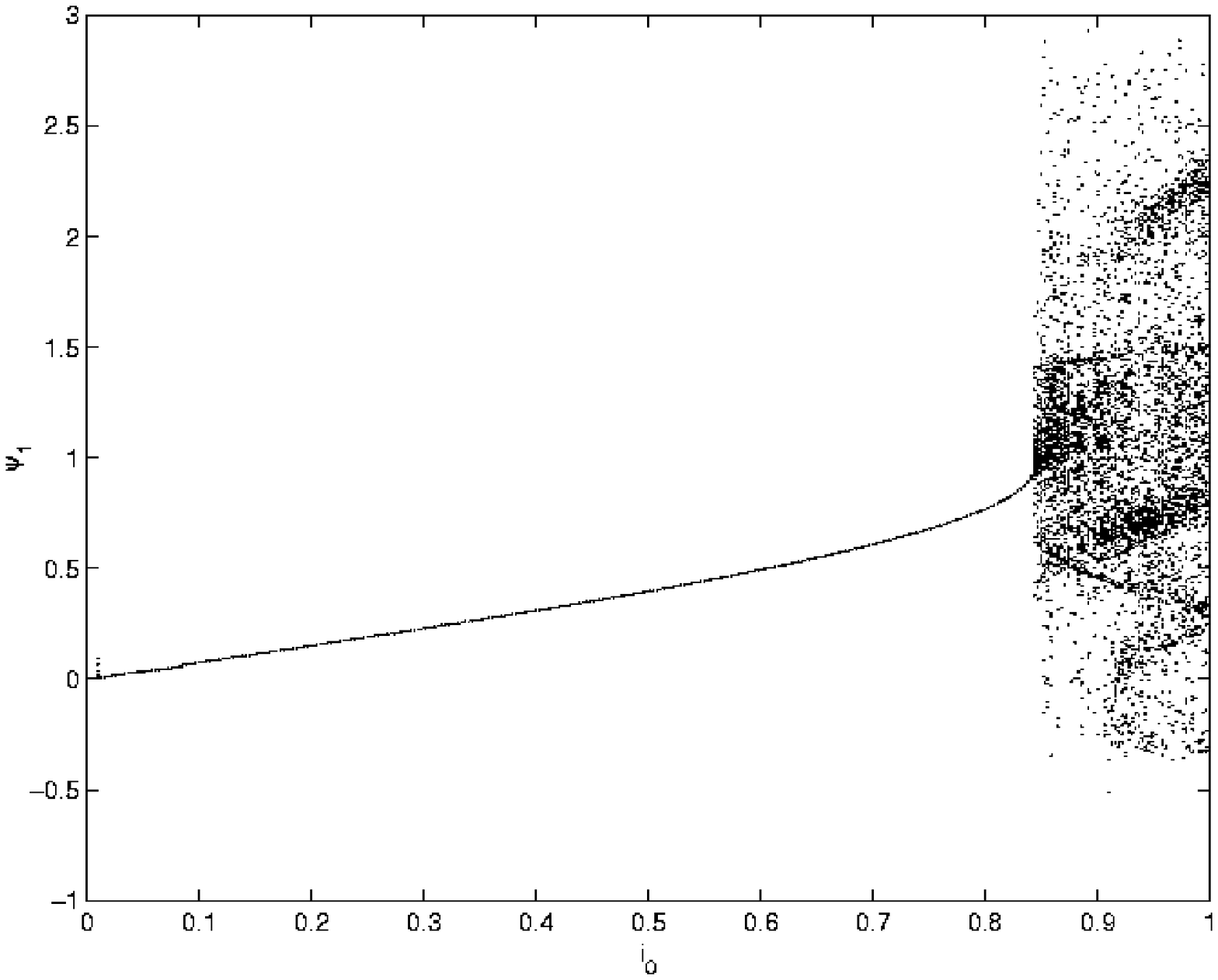}
\caption{Maxima of the normalized voltage is plotted against the
amplitude of applied field. It can be seen that the system is in
periodic motion for a wide range of amplitude values. $\theta=\pi$,
$\alpha_s=0.45$, $i_{dc}=0.3$,
$ \beta=0.15$ and $\omega=0.6$.} %
\label{volepsithe}
\end{figure}

Now we fix the value of $\alpha$ as $0.45$ and the amplitude of the
driving rf field is changed from $0$ to $1$. Without an applied
phase difference we can see from Fig.~\ref{volepsi} that the system
exhibits chaotic motion from a value of $0.43$ onwards with some
periodic windows in between. However on the application of a phase
difference the system stays in periodic state for a wide range of
amplitude values which were chaotic earlier. Fig.~\ref{volepsithe}
shows the response of the system when the amplitude of the driving
field is changed from $0$ to $1$ with an applied phase difference
of~$\pi$.
%%%%
%%%%%%%%%%%%%%%%555

\begin{figure}[tbh]
\centering
\includegraphics[width=0.8\columnwidth]{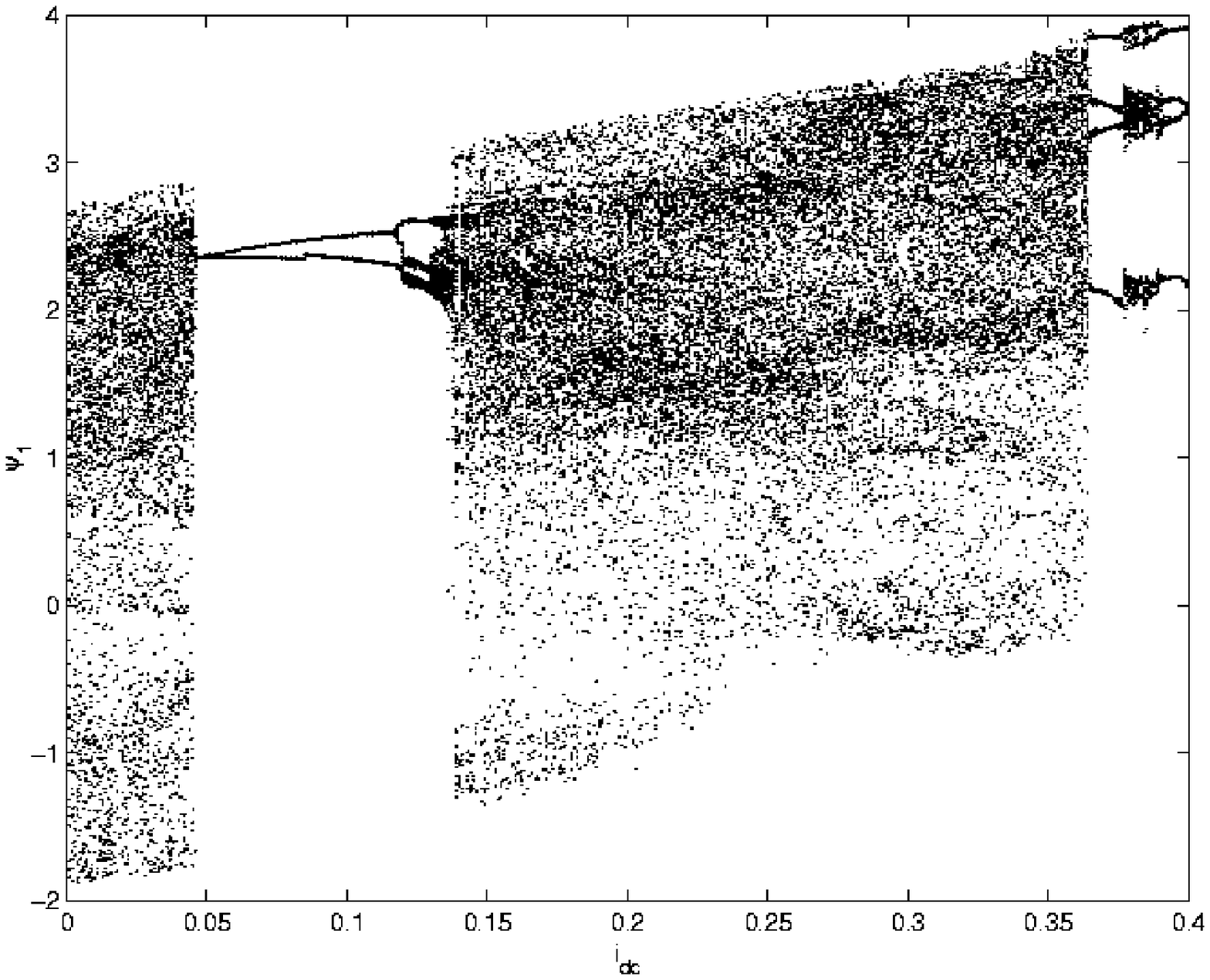}
\caption{Maxima of the normalized voltage against the $i_{dc}$.
$\theta=0$, $\alpha_s=0.45$,
$ \beta=0.15$, $i_0=0.7$ and $\omega=0.6$.} %
\label{volbet}
\end{figure}

\begin{figure}[tbh]
\centering
\includegraphics[width=0.8\columnwidth]{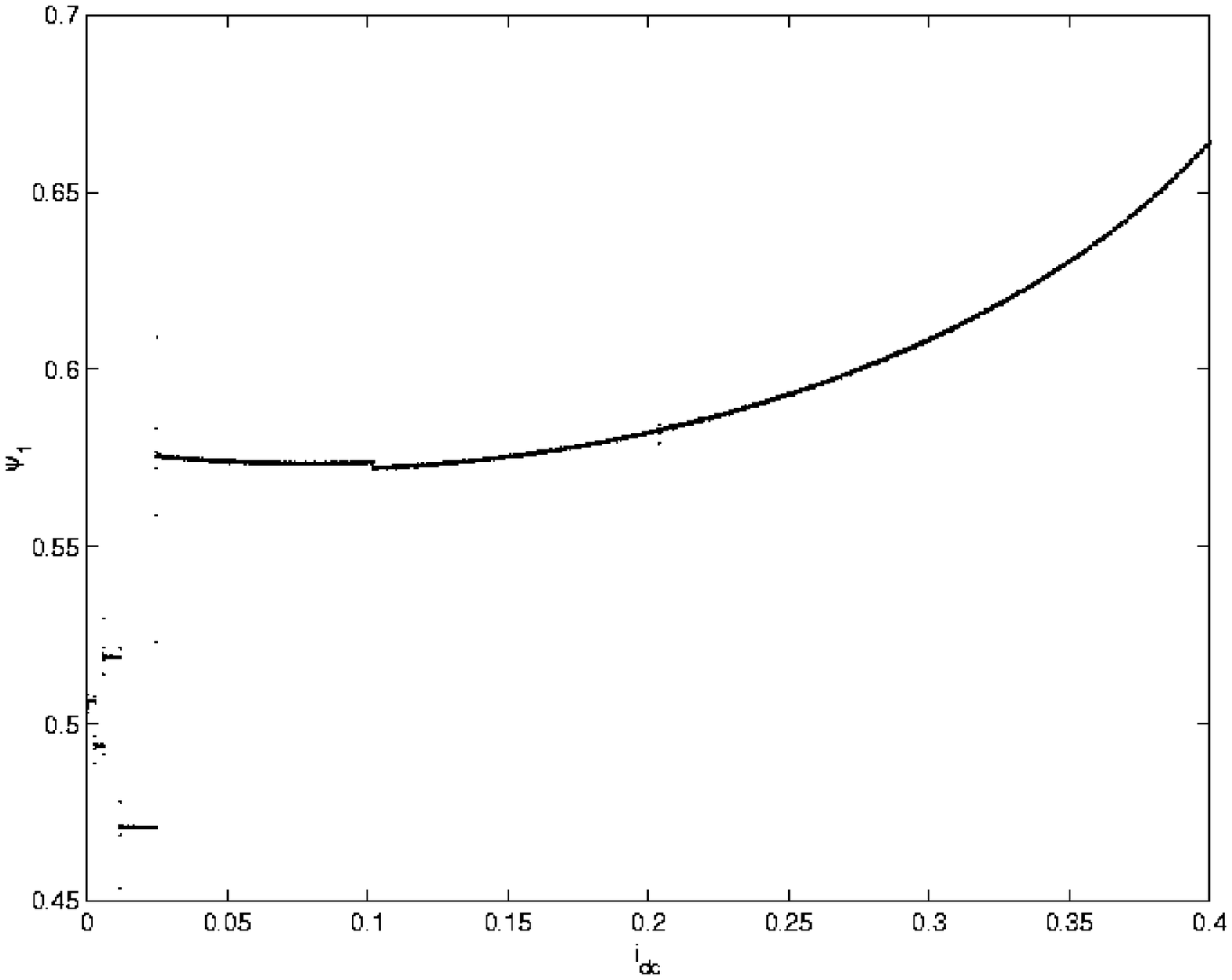}
\caption{Maxima of the normalized voltage against the $i_{dc}$ .
$\theta=\pi$, $\alpha_s=0.45$,
$ \beta=0.15$, $i_0=0.7$ and $\omega=0.6$.} %
\label{volbetthe}
\end{figure}

The effect of phase on the applied dc bias on the system is also
studied. For this all other parameter values were fixed and $i_{dc}$
value is changed from $0$ to $0.4$. Here also we observe that the
system continues to be in periodic motion for a large range of
$i_{dc}$ values. The comparison can be obtained from
Fig.~\ref{volbet} and Fig.~\ref{volbetthe}. Thus we show that the
system exhibits periodic motion for a wide range of parameter values
for an applied phase difference between the driving fields. This
might be of great practical importance when we consider Josephson
junction devices like voltage standards.

An important point to be noted is that the parameter values at which
we apply phase difference is to be chosen carefully. If the values
we choose is in  a region where the difference in voltage 
($\psi_1-\psi_2$) is large, then by just applying a phase difference
we may not be able to control chaos.
%%%%%%%%%
%%%%%%%%
\section{Conclusion}
In this work we find that by applying a phase difference in the
driving fields we can control chaos in Josephson junction. We also
observe that for this type of control we have to choose the junction
parameter values where the difference in voltages between the two
junctions is negligible. We report that in a region where the
difference in voltages between the two junction is not negligible
chaos cannot be controlled by applying a phase difference in the
driving fields. For the parameter region we choose the system was in
periodic motion for a phase difference of $\theta=0.95 \pi- 1.5
\pi$. Now we fix the phase difference as $\pi$ and vary other
parameters such as dc bias, amplitude of applied field and coupling
strength. It is found that even for large variation of these
parameters, the system continues to be in periodic motion. So this
might be of great practical importance as phase difference can be
easily applied to the rf-field in an experimental set up. Thus it
offers an easier way to control chaos and thus will provide an
enhanced capability to design superconducting circuits in such a way
as to maximize the advantages of non linearity while
minimizing the possibility of instabilities.%

\ack
One of the authors, C.R.N wishes to thank M.P.John, ISP, CUSAT for
the various fruitful discussions.

%%%%%%%%%%
%%%%
%\bibliographystyle{ncnsd}
%
\end{document}